\documentclass[pra,aps,twocolumn,preprintnumbers,amsmath,amssymb]{revtex4-1}
\usepackage{graphicx}
\usepackage{verbatim}
\newcommand{\ba}{\begin{array}}
\newcommand{\ea}{\end{array}}
\newcommand{\be}{\begin{equation}}
\newcommand{\ee}{\end{equation}}
\newcommand{\bea}{\begin{eqnarray}}
\newcommand{\eea}{\end{eqnarray}}
\begin{document}

\title{Models of semiconductor quantum dots blinking based on spectral diffusion }
\author{ Vladislav~K.~Busov}
\author{ Pavel~A.~Frantsuzov}
\email{frantsuzov@rector.msu.ru}
\affiliation{Lomonosov Moscow State University, 119991 Moscow, Russia}

\begin{abstract}
Three models of single colloidal quantum dot emission fluctuations (blinking) based on spectral diffusion were  considered analytically and numerically. It was shown that the only one of them, namely the Frantsuzov and Marcus model
reproduces the key properties of the phenomenon. The other two models,
the Diffusion-Controlled Electron Transfer (DCET) model  and the Extended DCET model predict that after an initial blinking period, most of the QDs should become permanently bright or  permanently dark which is significantly different from the experimentally observed behavior.
 \end{abstract}
\date{\today}
\maketitle

\section{Introduction}

Two decades have passed since the first observation of long-term fluorescence intensity fluctuations (blinking) of single colloidal CdSe quantum dots (QDs) with a ZnS shell \cite{BrusNature96}.  In further experimental studies it was found (see \cite{FrantsuzovNaturePhys08,BarkaiPT09,OrritCOCIS07,MulvaneyPCCP06,KraussJCPL10,ReidIJMS12,OronIJC12,LeoneCSR13} and references therein) that
these fluctuations have a wide spectrum of characteristic timescales,
from hundreds of microseconds to hours. The intensity traces (binned photon counting data) of CdSe/ZnS core/shell dots show the following key properties:\\
1. The intensity distribution usually has two maxima, so-called ON and OFF intensity levels\\
 2. The ON-time and OFF-time distributions obtained by the threshold procedure have the truncated power-law form\\
\be
p(t)\sim t^{-m} \exp(-t/T)
\label{tmexp}
\ee
3. The power spectral density of the trace has a $1/f^r$ dependence, where $r$ value is around 1. This dependence
changes to  $f^{-2}$ at large frequencies \cite{PeltonPNAS07}.

\begin{figure*}[ht]
\includegraphics[width=7 in]{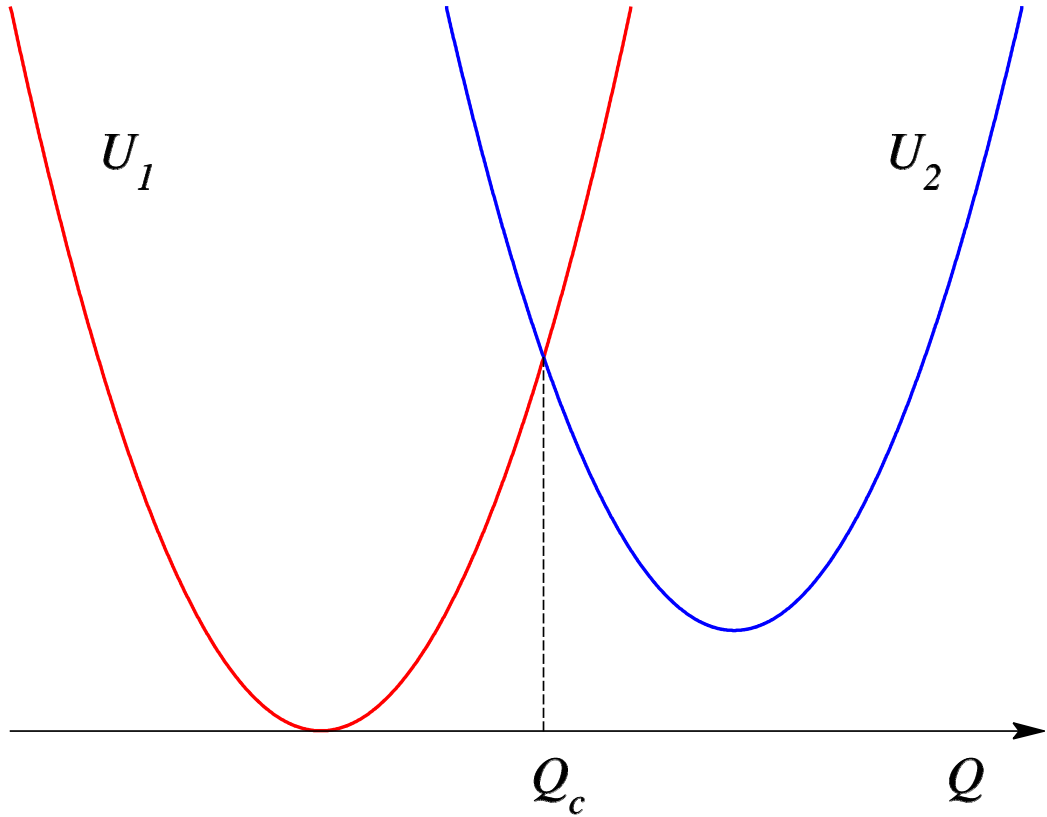}
\caption{The schematic picture of the DCET model.
The potential surfaces of the neutral (bright) and the charged (dark) electronic states are represented by
the red and blue lines, respectively.   Vertical dotted line corresponds to the crossing point.}
\label{fig:Tang}
\end{figure*}

Another interesting phenomenon that manifests in the emission of single quantum dots is the spectral diffusion
showing characteristic time scales in the order of hundreds of seconds \cite{BawendiPRL96,BawendiJPCB99}.
It is not surprising that there are a number of models proposed to explain the blinking that relate the fluctuations in the emission intensity with
slow variations in the exciton energy.
The first model of that kind suggested by Shimizu et al. \cite{BawendiPRB01}
is based on the Efros/Rosen {\it charging mechanism} (CM) \cite{EfrosPRL97}. The CM attributes the ON and OFF periods to neutral and charged QDs, respectively. The light-induced electronic excitation in the charged QD is supposed to be quenched by a fast Auger recombination process.
The model of Shimizu et al. \cite{BawendiPRB01} assumes that the charging/discharging events happen when the energies
of the neutral exciton and the charged state are in resonance.
A more advanced version of this idea was used by Tang and Marcus in the DCET model \cite{TangJCP05,TangPRL05}.
In 2014 Zhu and Marcus \cite{ZhuPCCP2014} presented an extension of the DCET model by introducing an additional biexciton charging channel.

 Simultaneously with Tang and Marcus \cite{TangJCP05,TangPRL05}, another diffusion model based on the alternative {\it fluctuating rate mechanism} (FRM) of blinking was suggested by Frantsuzov and Marcus \cite{FrantsuzovPRB05}.
 The FRM assumes that the non-radiative relaxation rate of the exciton is subject to long term fluctuations caused by the rearrangement of surface atoms.
A basic life cycle of the QD within this mechanism begins with a photon absorption. A relaxation of the excited state can go in one of of two paths. The first path is relaxation via a photon emission. The second path is a hole trapping followed by a consequent non-radiative recombination with a remaining electron. The photoluminescence quantum yield (PLQY) of the QD emission in this case can be expressed as
 \be
Y(t) = \frac{k_r}{k_r+k_t(t)}\equiv k_r \tau_{av}
\label{Y}
\ee
where $k_r$ is the radiative recombination rate, and $\tau_{av}$ is the averaged exciton lifetime.
Thus the variations of the $k_t$ generate fluctuations of the emission intensity on a long time scale.
The Frantsuzov and Marcus model \cite{FrantsuzovPRB05} connects the recombination rate with the fluctuating energy difference between 1S$_e$ and 1P$_e$ states.

In this article we are going to discuss the advantages and disadvantages of these models of single QD blinking based on spectral diffusion
as well as their perspectives of further development.

\section{Diffusion-controlled electron transfer model}

 After introducing the Marcus reaction coordinate $Q$, DCET model
 equations describing the evolution of its probability distribution density
 in the neutral state $\varrho_1(Q,t)$ and in the charged state $\varrho_2(Q,t)$ can be written in the following form:
$$\frac{\partial}{\partial t}\varrho_1(Q,t)= D_1 \frac{\partial}{\partial Q}
\left(\frac{\partial}{\partial Q}+ \frac{U_1'(Q)}{kT} \right) \varrho_1(Q,t)$$
\begin{equation}
-2\pi\frac {V^2} \hbar \delta(U_1(Q)-U_2(Q))\left(\varrho_{1}(Q,t)-\varrho_{2}(Q,t)\right)
\label{EqZ1}
\end{equation}
$$\frac{\partial}{\partial t}\varrho_2(Q,t)= D_2 \frac{\partial}{\partial Q}
\left(\frac{\partial}{\partial Q}+ \frac{U_2'(Q)}{kT}\right) \varrho_2(Q,t)$$
\begin{equation}
-2\pi\frac {V^2} \hbar \delta(U_1(Q)-U_2(Q))\left(\varrho_{2}(Q,t)-\varrho_{1}(Q,t)\right),
\label{EqZ2}
\end{equation}
where $D_1$ and $D_2$ are  diffusion coefficients in the neutral electronic state and charged state respectively, $V$ is the electronic coupling matrix element between the
neutral and charged states, and $T$ is the effective temperature.
The potential surfaces of the neutral $U_1(Q)$ and charged $U_2(Q)$ states are  Marcus' parabolas (see Fig. \ref{fig:Tang}):
\begin{equation}
U_1(Q)=\frac {(Q+E_r)^2} {4 E_r} \qquad  U_2(X)=\frac {(Q-E_r)^2}{4 E_r}+\Delta G
\label{U_Marcus}
\end{equation}
characterized by the reorganization energy $E_r$ and the free energy gap $\Delta G$.
Transitions between the neutral and charged states are determined by
the delta-functional sink in the crossing point $Q_c$ (local Golden rule), where $U_1(Q_c)=U_2(Q_c)$
$$Q_c=\Delta G$$

Equations (\ref{EqZ1}-\ref{EqZ2}) were initially introduced in 1980 independently by
Zusman \cite{ZusmanCP80} and Burshten and Yakobson \cite{BurshteinCP80} for describing solvent
effects in electron transfer reactions. In the literature they are usually called Zusman equations (see for example
the review article \cite{BarzykinACP02} and references therein).The rigorous derivation of the Eqs. (\ref{EqZ1}-\ref{EqZ2})
from the basic quantum level (Spin-Boson Hamiltonian) was made in Ref. \cite{FrantsuzovJCP99}.
The characteristic time scales of diffusion in the process of the electron transfer are of the order of picoseconds.
That is to say that the equations (\ref {EqZ1}-\ref{EqZ2}) were originally designed to work for completely different time scales.

The statistics of the ON  time blinking periods within the DCET model can be calculated
using the function $\rho_1(Q,t)$
which is a solution of the equation (\ref{EqZ1}) where the term describing the transfer
from the charged state to the neutral one is omitted:
$$\frac{\partial}{\partial t}\rho_1(Q,t)= D_1 \frac{\partial}{\partial Q}
\left(\frac{\partial}{\partial Q}+ \frac{U_1'(Q)}{kT} \right) \rho_1(Q,t)$$
\begin{equation}
-2\pi\frac {V^2} \hbar \delta\left(U_1(Q)-U_2(Q)\right)\rho_{1}(Q,t)
\label{EqZi1}
\end{equation}
with the initial condition describing the distribution function right after the transition from the charged state:
$$\rho_1(Q,0)=\delta(Q-Q_c)$$
The probability of the ON state being longer than $t$ (survival probability) is defined by the integral of the function $\rho_1(Q,t)$
\be
S_{\mbox {\tiny ON}}(t)=\int\limits_{-\infty}^\infty \rho_1(Q,t)\,dQ
\label{Sur}
\ee
The ON time distribution function is expressed as a derivative
\be
p_{\mbox {\tiny ON}}(t)=-\frac{d}{dt} S_{\mbox {\tiny ON}}(t)
\label{pON}
\ee

The analytical expression for the Laplace image of the ON time distribution function
$$ \tilde p_{\mbox {\tiny ON}}(s)=\int\limits_0^{\infty}  p_{\mbox {\tiny ON}}(t) e^{-st} \,dt$$
was found by Tang and Marcus \cite{TangJCP05,TangPRL05} (derivation details are given in Appendix A):
\be
\tilde p_{\mbox {\tiny ON}}(s)=\frac {W g_1(s)}{1+W g_1(s)}
\label{pONs}
\ee
where
\be
W=\frac {\sqrt {2 \pi} V^2}{\hbar \sqrt{E_rkT}}
\label{W}
\ee
Function $g_1(s)$ can be expressed as an integral
\begin{equation}
g_1(s)=\int\limits_0^\infty \frac{\exp\left[-st-\frac {x_c^2}2\tanh\left({\frac{t}{2\tau_1}}\right)\right] }{\sqrt{2\pi\left(1-e^{-2t/\tau_1}\right)}}\,dt
\label{gs}
\end{equation}
where $\tau_1$ is the relaxation time in the the neutral state
\be
\tau_1=\frac {2E_rkT}{D_1}
\label{tau1}
\ee
and $x_c$ is the dimensionless crossing point coordinate
\be
x_c=\frac {E_r+\Delta G} {\sqrt {2E_rkT}}
\label{xc}
\ee

At a short time limit $t\ll \tau_1$ Tang and Marcus \cite{TangJCP05,TangPRL05}
presented the following approximation for the ON time distribution (see Appendix B):
\be
p_{\mbox {\tiny ON}}(t)=\frac{\exp(-\Gamma_1 t)}{\sqrt{\pi t_c t}} \left[1-\sqrt{\frac{\pi t}{t_c}} \exp\left(\frac t {t_c} \right)\mbox{erfc}\left(\sqrt{\frac t {t_c}}\right) \right]
\label{pONshort}
\ee
where
\be
\Gamma_1=\frac {x_c^2}{4\tau_1}
\label{Gamma1}
\ee
and $t_c$ is the critical time
\be
t_c=\frac 4 {W^2 \tau_1}
\label{tc}
\ee

When $t$ is much shorter than the critical time   Eq.(\ref{pONshort})  can be approximated as
\be
p_{\mbox {\tiny ON}}(t)\approx \frac 1 {\sqrt{\pi t_c}} t^{-1/2} , \quad t\ll t_c
\label{Ponshort}
\ee
when for longer times
\be
p_{\mbox {\tiny ON}}(t)\approx \frac 1 2 \sqrt{\frac {t_c} \pi} t^{-3/2}\exp(-\Gamma_1 t), \quad  t_c\ll t \ll \tau_1
\label{Ponlong}
\ee
The equation (\ref{Ponlong}) reproduces the experimentally observed truncated power-law dependence Eq. (\ref{tmexp}).
This dependence has to correspond to the power spectral density of the emission intensity $S(f)\sim f^{-3/2}$.
The experimentally observed transition of the power spectral density dependence
 to $f^{-2}$ at large frequencies \cite{PeltonPNAS07} was explained by the changing of the ON time distribution
 function behavior from (\ref{Ponshort}) to (\ref{Ponlong}) at times $t\sim t_c$.

\begin{figure*}[ht]
\includegraphics[width=7 in]{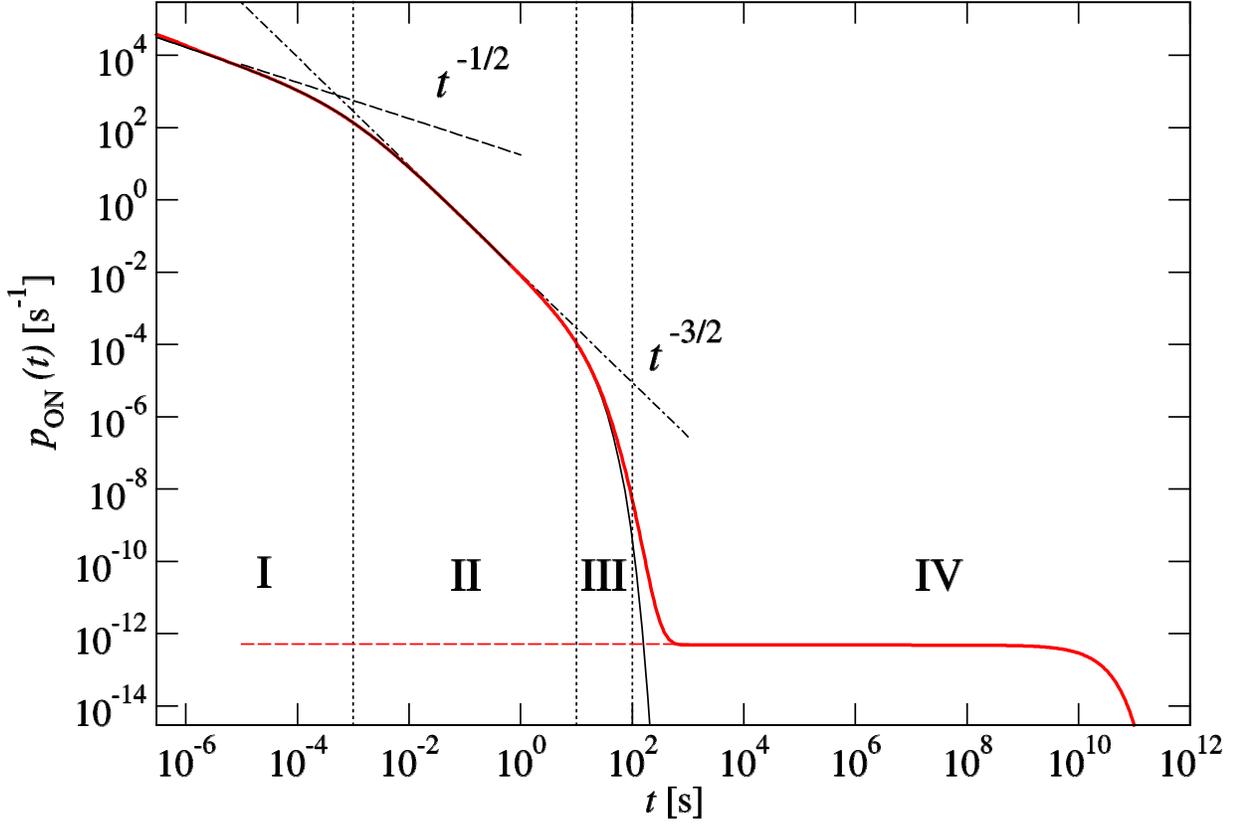}
\caption{The ON time distribution function within the DCET model (thick red line),  the first interval power law
(black dashed line), the second interval power law (black dashed-dotted line), the Tang-Marcus approximation Eq.(\ref{pONshort}) (thin black line) and the
long-time asymptotic Eq.(\ref{Pexp}) (red dashed line). Vertical dotted lines represent borders between characteristic intervals at $t_c$, $1/\Gamma_1$ and $\tau$.
The parameters of the model are  $\tau_1=100\,s$, $\Gamma_1=0.1\,s^{-1}$, $t_c=10^{-3}\,s$.}
\label{fig:f1}
\end{figure*}

\begin{figure*}[ht]
\includegraphics[width=7 in]{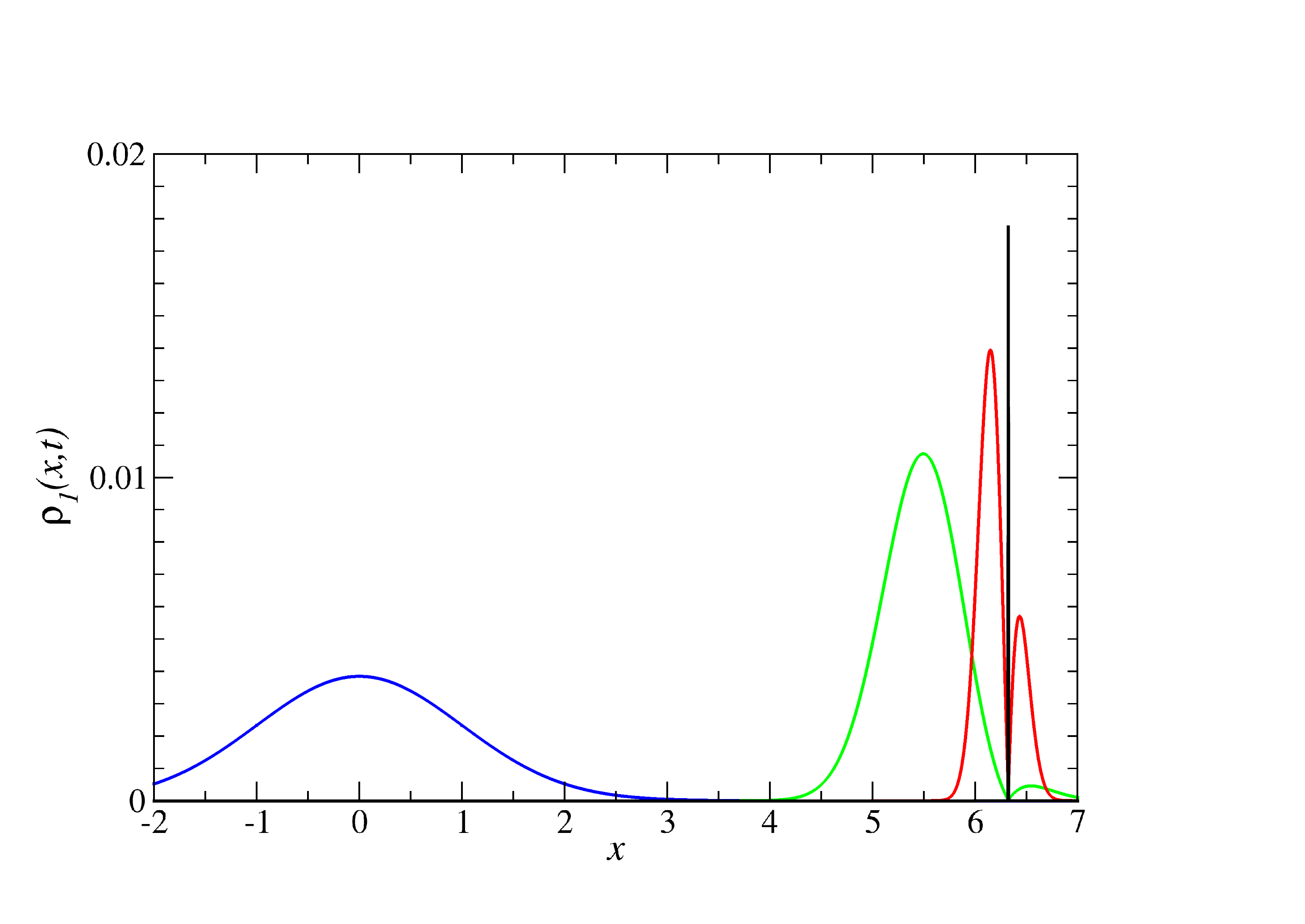}
\caption{The coordinate probability distribution function $\rho_1(x,t=10^{-4}\,s)\times 10^{-4}$ (black line),
 $\rho_1(x,t=1\,s)\times 0.25$ (red line), $\rho_1(x,t=10.21\,s)$ (green line), $\rho_1(x,t=10^{5}\,s)$ (blue line).
The parameters of the model are  $\tau_1=100\,s$, $\Gamma_1=0.1\,s^{-1}$, $t_c=10^{-3}\,s$. $x$ is the dimensionless coordinate
 $x=(Q+E_r)/\sqrt {2E_rkT}$}
\label{fig:f2}
\end{figure*}

\begin{figure*}[ht]
\includegraphics[width=7 in]{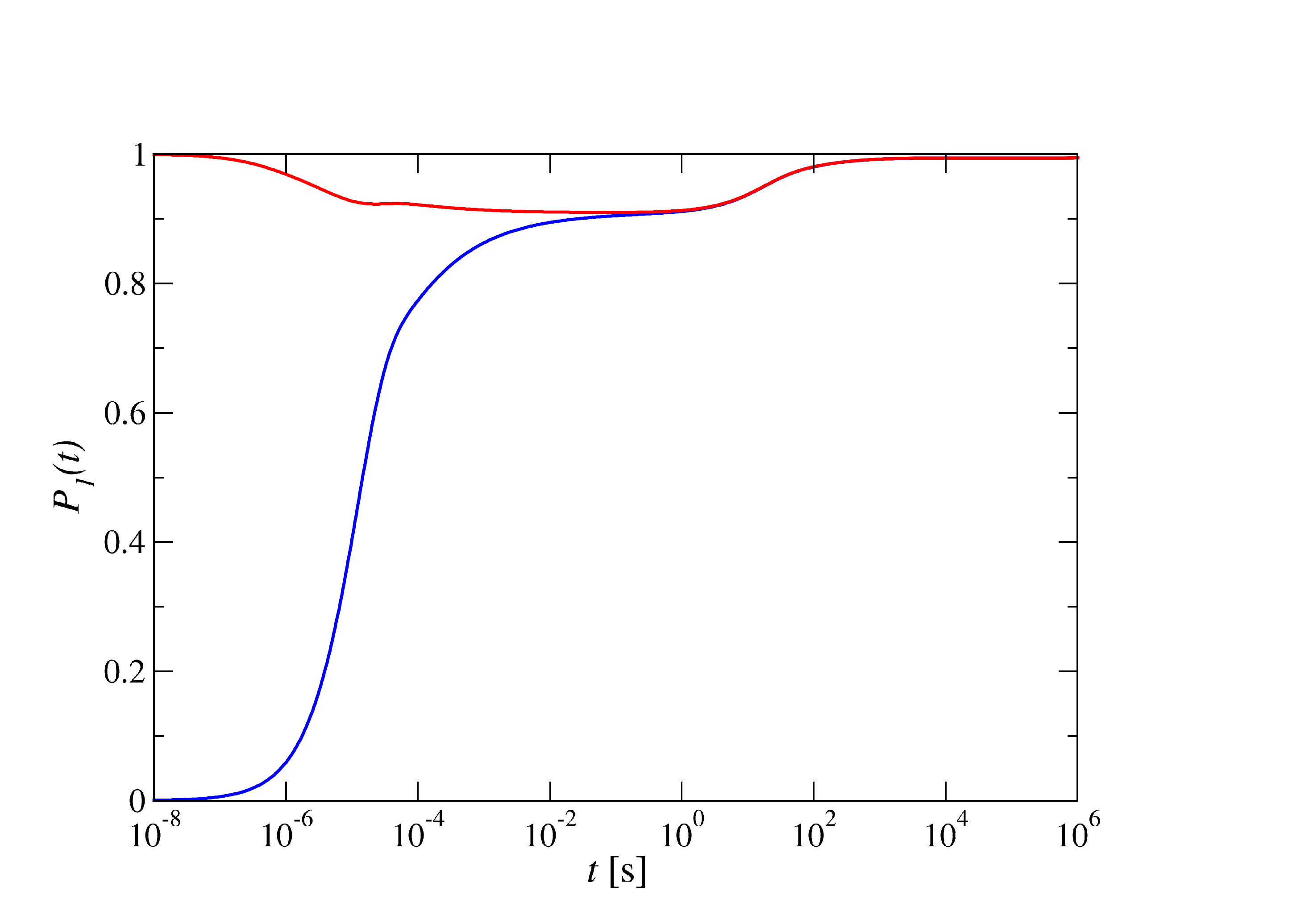}
\caption{The time dependence of the probability of finding the QD in the neutral (bright) state for the initial condition (\ref{CondON})
 (red line) and the initial condition (\ref{CondOFF}) (blue line).
The parameters of the model are  $\tau_1=100\,s$, $\Gamma_1=0.1\,s^{-1}$, $t_c=10^{-3}\,s$, $\tau_2=10^4\,s$, $\Gamma_2=10^{-3}\,s^{-1}$}
\label{fig:f3}
\end{figure*}

The problem is that for longer times $t \gg \tau_1$ the approximate formula (\ref{pONshort}) is not applicable.
It can be shown (see Appendix C)  that at a very long time scale the ON time distribution shows
slow exponential decay \cite{TangPRL05}:
\be
  p_{\mbox {\tiny ON}}(t)\approx p_1 \exp(-k_1t) , \quad \tau_1 \ll t
  \label{Pexp}
\ee
were $k_1$ is the decay rate
\be
k_1=\frac {W}{\sqrt{2\pi}(1+WB)}\exp\left(-\frac {x_c^2}2\right)
\label{k1}
\ee
 $p_1$ is the amplitude
\be
p_1=\frac {k_1} {1+WB}
\label{p1}
\ee
and
$$B=\int\limits_0^\infty \left[\frac{\exp\left(-\frac {x_c^2} 2 \tanh\left({\frac{t}{2\tau_1}}\right)\right) }
 {\sqrt{2\pi\left(1-e^{-2t/\tau_1}\right)}}-\frac{\exp\left(-\frac {x_c^2}2\right)}{\sqrt{2\pi}}\right]\,dt$$
The last integral can be expressed in terms of a generalized hypergeometric function $_2F_2$ \cite{ZharikovJCP92}:
 \begin{equation}
B=\frac{\tau_1}{\sqrt{2\pi}} \exp\left(-\frac {x_c^2}2\right)\left[
\ln 2+x_c^2\,{_2F_2}\left(\left.\begin{array}{cc}1&1\\\frac{3}{2}&2\end{array}
\right|\frac{x_c^2}{2}\right)\right]
\label{B}
\end{equation}
 The simpler analytical expressions  of $B$  can be found in the limiting cases \cite{ZharikovJCP92}:
 \be
 B\approx \left\{ \begin{array} {ll}\tau_1 {\ln 2}/\sqrt {2 \pi},&  \quad |x_c|\ll 1\\
                                     \tau_1/{|x_c|} , & \quad |x_c|\gg 1\end{array} \right.
 \label{Bapprox}
\ee

Equation (\ref{k1}) can be rewritten as
\be
k_1=\frac {W}{\sqrt{2\pi}(1+WB)}\exp\left(-\frac{(E_r+\Delta G)^2}{4E_rkT}\right)
\label{k1M}
\ee
This formula is well-known in electron transfer theory \cite{BarzykinACP02}.
It describes the quasi-stationary  rate of the electron transfer in the absence of back transitions.
The argument in the exponent  reproduces the famous Marcus' Free Energy Gap law.
For low coupling values the rate Eq.(\ref{k1M}) is proportional to $V^2$  (the  Golden Rule result):
 $$ k_1= \frac {V^2}{\hbar \sqrt{E_rkT}}\exp\left(-\frac{(E_r+\Delta G)^2}{4E_rkT}\right)$$
 At high coupling values the rate is limited by the diffusion transport to the crossing point and so becomes independent of $V$.
 For the activated process   $(E_r+\Delta G)^2\gg 4E_rkT$  from  Eqs.(\ref{k1M}) and (\ref{Bapprox}) we get:
  $$k_1=\frac {|E_r+\Delta G|}{\tau_1 \sqrt{4\pi E_rkT}  }\exp\left(-\frac{(E_r+\Delta G)^2}{4E_rkT}\right)$$
 The maximum rate  is reached in the activationless case $(E_r+\Delta G)^2\ll 4E_rkT$
 $$ k_1= \frac {1}{\tau_1 \ln 2}$$
 As we can see the rate $k_1$ is always less than $1/\tau_1$.

 The OFF time distribution shows a similar behaviour:
$$p_{\mbox {\tiny OFF}}(t)\approx \frac 1 {\sqrt{\pi t_2}} t^{-1/2}, \quad t\ll t_2 $$
$$p_{\mbox {\tiny OFF}}(t)\approx \frac 1 2 \sqrt{\frac {t_2} \pi} t^{-3/2}\exp(-\Gamma_2 t), \quad  t_2\ll t \ll \tau $$
$$p_{\mbox {\tiny OFF}}(t)\approx p_2 \exp(-k_2t), \quad \tau_2 \ll t$$
where
$$\Gamma_2=\frac {x_2^2}{4\tau_2},\quad t_2=\frac 4 {W^2 \tau_2}$$
$$k_2=\frac {W}{1+WB_2}\exp\left(-\frac {x_2^2}2\right), \quad p_2=\frac {k_2} {\sqrt{2\pi}(1+WB_2)}$$
and
\begin{equation}
B_2=\frac{\tau_2}{\sqrt{2\pi}}\exp\left(-\frac {x_2^2}2\right)\left[
\ln 2+x_2^2\,{_2F_2}\left(\left.\begin{array}{cc}1&1\\\frac{3}{2}&2\end{array}
\right|\frac{x_2^2}{2}\right)\right]
\end{equation}

 According to Eq.(\ref{pONshort}) and Eq. (\ref{Pexp}) there are four characteristic time intervals of the $p_{\mbox {\tiny ON}}(t)$ behavior:\\
{\bf Interval I:}  Power-law  with $1/2$ exponent at $t\ll t_c$;\\
{\bf Interval II:} Power-law  with $3/2$ exponent at $t_c\ll t \ll 1/\Gamma_1$;\\
{\bf Interval III:} Exponential decay  at $1/\Gamma_1 \ll t \ll \tau_1$;\\
{\bf Interval IV:} Long time exponential decay  $\tau_1 \ll t$.\\
Note that Interval III can only exist if
\be
\Gamma_1\tau_1\gg 1
\label{Gtau}
\ee

We performed numerical simulations of Eq.(\ref{EqZi1}) using the SSDP program \cite{KrissinelJCC97}.
The results of the simulations for the parameters $\tau_1=100\,s$, $\Gamma_1=0.1\,s^{-1}$, $t_c=10^{-3}\,s$ are presented in Fig. \ref{fig:f1}.
The parameters are very close to the ones used in Ref.\cite{TangJCP05} for fitting the experimental data.
The model parameters can be restored using  Eqs.(\ref{Gamma1}) and (\ref{W}):
 $$x_c=\sqrt{4\Gamma_1\tau_1}\approx 6.32,\quad W=\sqrt{\frac 4 {\tau_1 t_c}}\approx 6.32\, s^{-1}$$
 The condition $x_c\gg 1$ following from (\ref{Gtau})is satisfied.
  Using Eq.(\ref{Bapprox}) we get
  $$BW=\frac 1 {\sqrt{\Gamma_1 t_c}}=100$$
 An expression for $k_1$ follows from  Eq.(\ref{k1})
 $$k_1= \sqrt{\frac {2 \Gamma_1} {\pi\tau_1}} \frac { \exp(-2 \Gamma_1 \tau_1)} {1+\sqrt{\Gamma_1 t_c}}\approx 5.15\times 10^{-11}\, s^{-1} $$
All four characteristic intervals of the ON time distribution dynamics are clearly seen on Fig. \ref{fig:f1}.

The value $ p_{\mbox {\tiny ON}} (t) $ is very small at $t \gg \tau_1$ (interval IV), however
the probability for the ON state to survive after $\tau_1$ time is quite significant.
$$S_1= S_{\mbox {\tiny ON}}(t \gg \tau_1) \approx \int\limits_{0}^\infty p_1 \exp(-k_1 t)\,dt$$
From (\ref{p1}) we get
$$S_1\approx \frac 1 {1+BW}=\frac {\sqrt{\Gamma_1 t_c}} {1+ \sqrt{\Gamma_1 t_c}} \approx 0.01$$
That is why the averaged ON time is extremely long:
 $$\langle t_{\mbox {\tiny ON}}\rangle=\int\limits_{0}^\infty t p_{\mbox {\tiny ON}}(t)\,dt \approx \int\limits_{0}^\infty t p_1\exp(-k_1 t)\,dt$$
and after integration:
$$\langle t_{\mbox {\tiny ON}} \rangle\approx \frac {k_1^{-1}} {1+BW}=\sqrt{\frac{\tau_1 t_c}{2}}\exp(2\Gamma_1 \tau_1)\approx 1.08\times 10^8 s$$

The coordinate probability distribution function $\rho_1(Q,t)$
within each interval is shown on Fig. \ref{fig:f2}.
At a short time (Interval I) the distribution has one narrow maximum, its width increases with time
 $\Delta Q = \sqrt{2 D_1t}$. The distribution function value at the crossing point
 $\rho_1(Q_c,t)$ decays as $\sim t^{-1/2}$ and it follows the same power law form
 of the ON time distribution.
 At longer times (Interval II) the delta-functional sink burns a hole in
 the distribution function,  and it shows two maxima.
 The distribution starts shifting towards the potential minimum within Interval III.
 That shifting generates an exponential decreasing of the  $\rho_1(Q_c,t)$ and
 as a result the exponential decay of the ON time distribution function.
 At times longer than $\tau_1$ (Interval IV) the function  $\rho_1(Q,t)$
reaches  the quasistationary distribution at the bottom of the parabolic potential
$$\rho_1(Q,t) \approx \frac{S_1}{\sqrt{4\pi E_rkT}} \exp\left(-\frac {(Q+E_r)^2}{4E_rkT}\right)\exp(-k_1t)$$
As such, the transition to the OFF state can only occur at the $Q_c$  crossing point, which requires thermal activation.
This explains why the decay of the ON time distribution is so gradual within Interval IV.

As seen from the analytical analysis and numerical simulations  the DCET model predicts the appearance of extremely long ON time periods
in a single QD emission trace. As seen on Fig. \ref{fig:f1} such a period could last years, which is much longer than the duration of a typical experiment.
The probability of such a long duration of a single ON time blinking event $S_1$ is found to be in order of  1\%.
Thus the QD can become permanently bright after about one hundred blinking cycles with a high probability.

All the predictions made about the ON time distribution can be applied  for the OFF distribution as well.
In most experiments the OFF time distribution truncation time of the single QD emission trace is too
long to be detected.  The only exceptions are the observations made on similar nanoobjects, namely  nanorods
 \cite{DrndicNL08}  where the value $1/\Gamma_2\sim 2500\,s$ was found.
Let us set  $\Gamma_2=  10^{-3}\,s^{-1}$ and $\tau_2=10^4\,s$.
The corresponding rate for long time decay is $k_2\approx 5.15\times 10^{-13}\, s^{-1}$
 The probability of an extremely long OFF time period is
$$S_2= \frac {\sqrt{\Gamma_2 t_2}} {1+ \sqrt{\Gamma_2 t_2}} \approx  10^{-4}$$
This means that after  about ten thousand blinking cycles the QD should
become permanently bright or permanently dark.
This prediction significantly differs from the behavior of single quantum dots observed in numerous experiments.

The fact that $S_2$ is much smaller than $S_1$ ($S_2/S_1 \approx 10^{-2}$) suggests
that the most of the QDs should became permanently bright.
In order to verify that statement we used the SSDP program \cite{KrissinelJCC97} for numerical simulations
of the Eqs.(\ref{EqZ1}-\ref{EqZ2}) with two types of initial conditions:
at the beginning of the ON time period (delta-functional distribution in the neutral state)
\be
\varrho_1(Q,0)=\delta(Q-Q_c);\quad \varrho_2(Q,0)=0
\label{CondON}
\ee
and at the beginning of the OFF time period
\be
\varrho_1(Q,0)=0;\quad \varrho_2(Q,0)=\delta(Q-Q_c)
\label{CondOFF}
\ee

As shown in Fig. \ref{fig:f3}, the probability of finding the system in the ON state
$$ P_1 (t) = \int\limits_{-\infty}^\infty \varrho_1 (Q, t) \, dQ $$
becomes very close to unity at times greater than 100 seconds for both cases.

\section{Extended DCET model}

\begin{figure*}[ht]
\includegraphics[width=7 in]{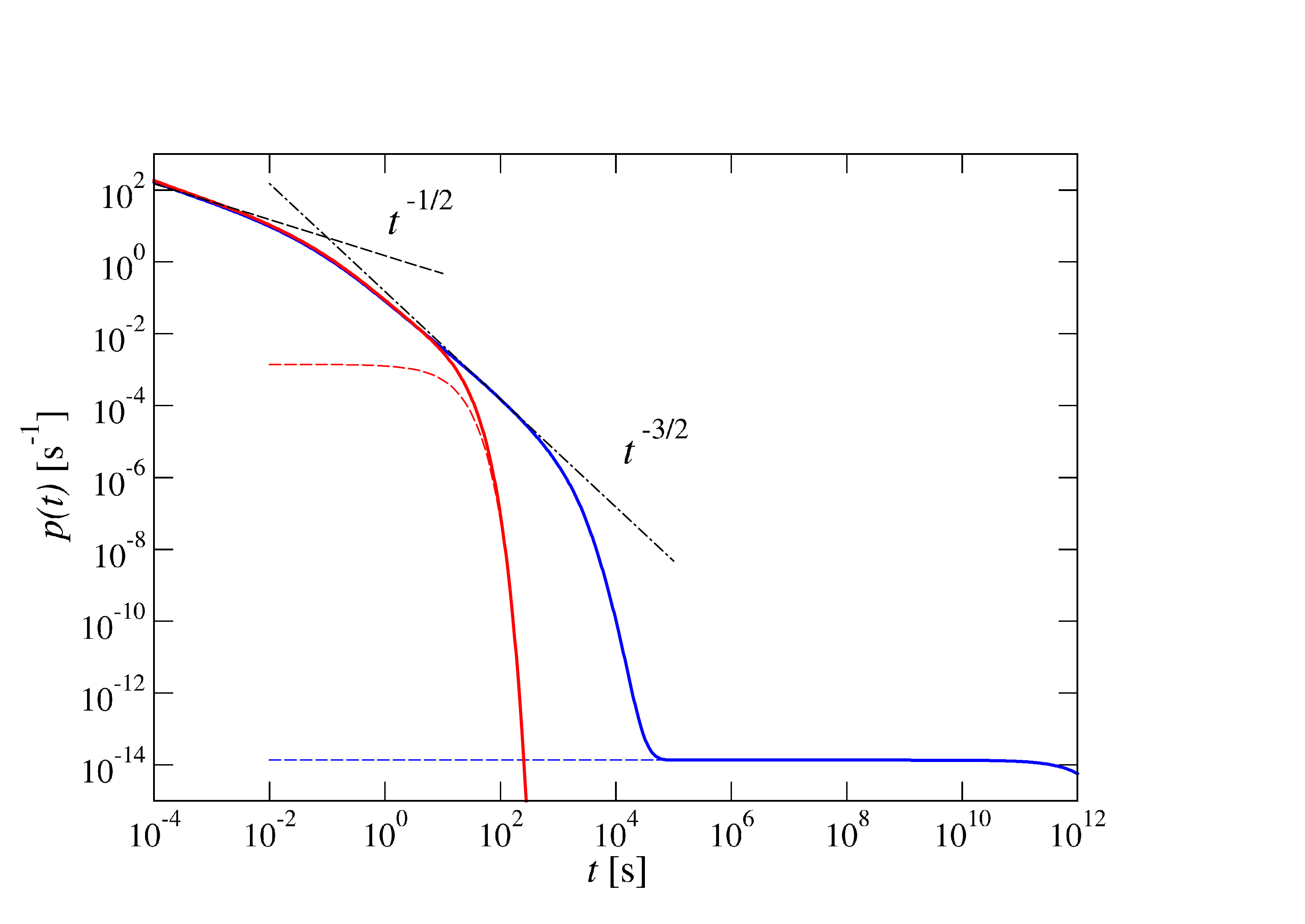}
\caption{The ON time distribution function (thick red line) and the OFF time distribution function (thick blue  line)  within the Extended DCET model,
the first interval power law (black dashed line), the second interval power law (black dashed-dotted line), the exponential decay of the ON time distribution Eq.(\ref{pONshort}) (dashed red line) and the OFF time distribution
long-time asymptotic Eq.(\ref{Pexp}) (blue dashed line).
The parameters of the model are  $\tau_1=\tau_2=10^4\,s$, $\Gamma_1=\Gamma_2=10^{-3}\,s^{-1}$, $t_c=t_2=0.1\,s$, $K_{L}=10^{-1}\,s^{-1}$}
\label{fig:f4}
\end{figure*}

\begin{figure*}[ht]
\includegraphics[width=7 in]{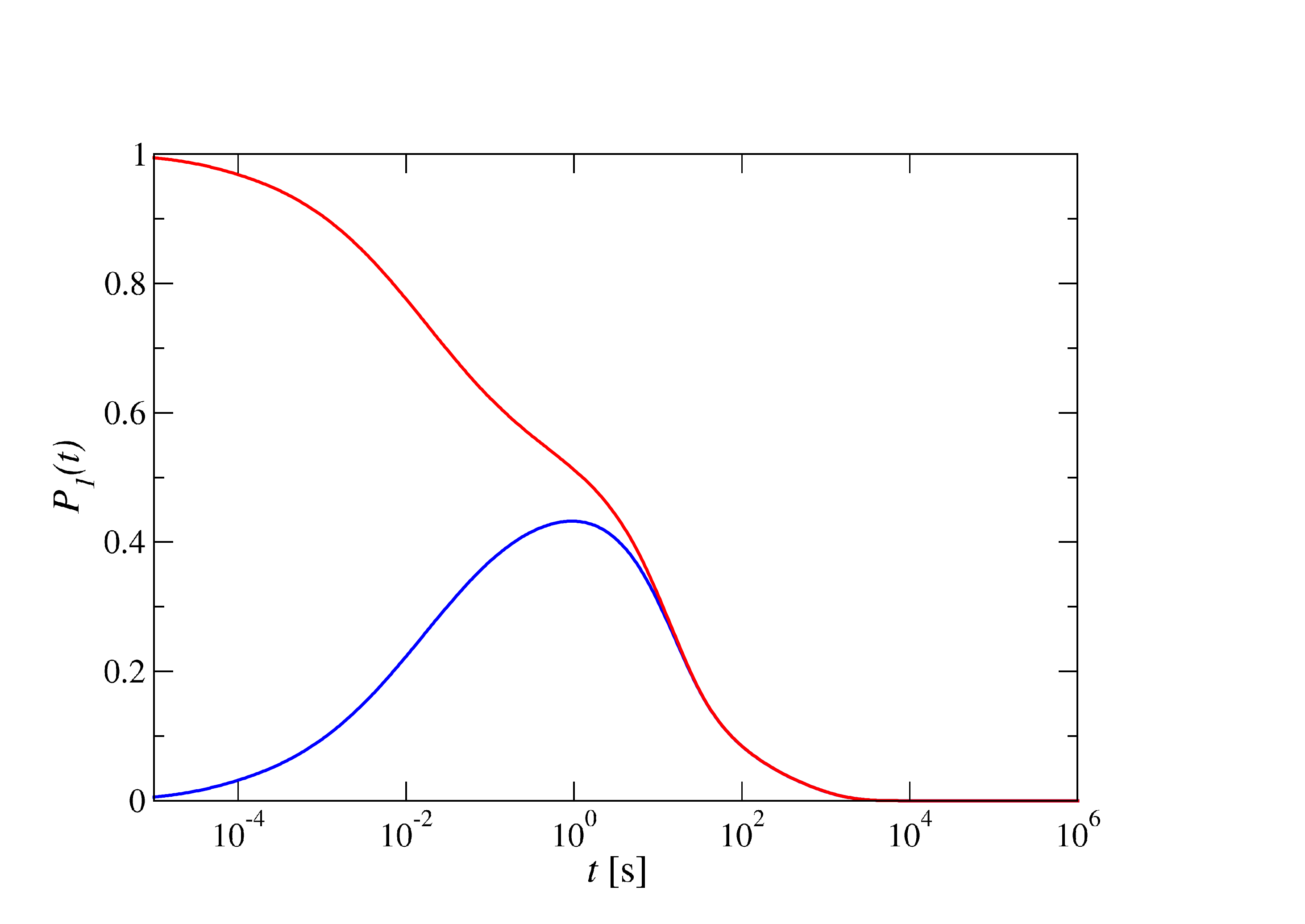}
\caption{The time dependence of the probability of finding the QD in the neutral (bright) state for the initial condition (\ref{CondON})
 (red line) and the initial condition (\ref{CondOFF}) (blue line) within the Extended DCET model.
 The parameters of the model are  $\tau_1=\tau_2=10^4\,s$, $\Gamma_1=\Gamma_2=10^{-3}\,s^{-1}$, $t_c=t_2=0.1\,s$, $k_{L}=10^{-1}\,s^{-1}$}
\label{fig:f5}
\end{figure*}

The extended DCET model of Zhu and Marcus \cite{ZhuPCCP2014} includes the
equations describing the evolution of the probability  density
  of the ground state $\varrho_g(Q,t)$,  the excited state $\varrho_e(Q,t)$, the biexciton state $\varrho_b(Q,t)$, the
charged (dark) state $\varrho_d(Q,t)$,  and the excited dark state $\varrho_{d^\ast}(Q,t)$:
\be
\frac{\partial}{\partial t}\varrho_g(Q,t)=k_{eg}\varrho_e(Q,t)-I_{ge}\varrho_g(Q,t)
\label{Zhu1}
\ee
$$
\frac{\partial}{\partial t}\varrho_e(Q,t)=I_{ge}\varrho_g(Q,t)+L_e\varrho_e(Q,t)+k_{be}\varrho_b(Q,t)$$
\be
-(k_{eg}+I_{eb})\varrho_e(Q,t)-k_{ed}\delta(Q-Q_c)\varrho_e(Q,t)
\label{Zhu2}
\ee
\be
\frac{\partial}{\partial t}\varrho_b(Q,t)=I_{eb}\varrho_e(Q,t)+L_b\varrho_b(Q,t)-(k_{be}+k_{bd'})\varrho_b(Q,t)
\label{Zhu3}
\ee
\be
\frac{\partial}{\partial t}\varrho_d(Q,t)=k_{d^\ast d}\varrho_{d^\ast}(Q,t)-I_{dd^\ast}\varrho_d(Q,t)
\label{Zhu4}
\ee
$$ \frac{\partial}{\partial t}\varrho_{d^\ast}(Q,t)=L_{d^\ast}\varrho_{d^\ast}(Q,t)+I_{dd^\ast}\varrho_d(Q,t)$$
\be
-k_{d^\ast d}\varrho_{d^\ast}(Q,t)-k_{d^\ast e}\delta(Q-Q_c)\varrho_{d^\ast}(Q,t)
\label{Zhu5}
\ee
where $L_e$, $L_b$ and $L_{d^\ast}$ are diffusion operators
$$L_e= D_e \frac{\partial}{\partial Q} \left(\frac{\partial}{\partial Q}+ \frac{U_e'(Q)}{kT} \right)$$
$$L_b= D_b \frac{\partial}{\partial Q} \left(\frac{\partial}{\partial Q}+ \frac{U_b'(Q)}{kT} \right)$$
$$L_{d^\ast}= D_{d^\ast} \frac{\partial}{\partial Q} \left(\frac{\partial}{\partial Q}+ \frac{U_{d^\ast}'(Q)}{kT} \right)$$
$D_e$, $D_b$ $D_{d^\ast}$ are the diffusion coefficients, $U_e(Q)$, $U_b(Q)$ and  $U_{d^\ast}(Q)$ are the potential surfaces of the excited state, the biexciton state and the dark excited state, respectively.
$ I_{ge}$,   $I_{eb}$, $I_{dd^\ast}$, $k_{eg}$, $k_{be}$, $k_{bd'}$, $k_{d^\ast d}$, $k_{d^\ast e}$, and
$k_{ed}$ are the rate constants.

The equation for the probability density of the higher energy dark state has to be added to the equation system (\ref{Zhu1}-\ref{Zhu5}):
\be
\frac{\partial}{\partial t}\varrho_{d'}(Q,t)=k_{bd'}\varrho_{b}(Q,t)-k_{d'd}\varrho_{d^\ast}(Q,t)
\label{Zhu6}
\ee

As stated by Zhu and Marcus \cite{ZhuPCCP2014} quasiequilibrium is established between the ground, the excited state and the biexciton state.
We can see from Eq. (\ref{Zhu6}) that a quasistationary  distribution of the the higher energy dark state is also determined by $\varrho_e (Q,t)$ and
it can also can be considered a part of the  quasiequilibrium. As such we can introduce the population of the integrated ON state
\be
\varrho_1 (Q,t)=\varrho_g (Q,t)+\varrho_e (Q,t)+\varrho_b (Q,t)+\varrho_{d'} (Q,t)
\label{Zhu7}
\ee
Similarly, there is a quasiequilibrium between the dark and the excited dark  states
and the OFF state population can also be introduced
\be
\varrho_2 (Q,t)=\varrho_d (Q,t)+\varrho_{d^\ast} (Q,t)
\label{Zhu8}
\ee
The following kinetic equations for the functions $\varrho_1(Q,t)$ and $\varrho_2 (Q,t)$ were obtained from Eqs. (\ref{Zhu1}-\ref{Zhu6})
(see Appendix D):
$$
\frac{\partial}{\partial t}\varrho_1(Q,t)=L_1\varrho_1(Q,t)-k_{L}\varrho_1(Q,t)$$
\be
-W_1\delta(Q-Q_c)\varrho_1(Q,t)  +W_2\delta(Q-Q_c)\varrho_2(Q,t)
\label{rhoI}
\ee
$$
\frac{\partial}{\partial t}\varrho_2(Q,t)=L_2\varrho_1(Q,t)-W_2\delta(Q-Q_c)\varrho_2(Q,t)$$
\be
+W_1\delta(Q-Q_c)\varrho_1(Q,t) +k_{LD}\varrho_1(Q,t)
\label{rhoII}
\ee
where $L_1$ and $L_2$ are effective diffusion operators:
$$ L_1=C_1\left(L_e+\frac {I_{eb}} {k_{be}}L_b\right); \quad L_2=C_2 L_d$$
$W_1$, $W_2$ and $k_{L}$ are effective rates:
$$W_1=C_1 k_{ed}; \quad W_2= C_2 k_{d^\ast e}; \quad k_{L} =C_1 k_{bd'}  \frac {I_{eb}} {k_{be}}$$
and $C_1$ and $C_2$ are the coefficients:
$$ C_1=\left( 1+\frac {k_{eg}} {I_{ge}} +\frac {I_{eb}} {k_{be}} +\frac {k_{bd'}} {k_{d'd}} \frac {I_{eb}} {k_{be}}\right)^{-1};\quad C_2=\left(1+\frac {k_{d^\ast d}} {I_{ge}}\right)^{-1}$$
We have to note that the equations  derived by  Zhu and Marcus ( Eqs.(11-12) in Ref.\cite{ZhuPCCP2014}) using the same procedure
are different from Eqs. (\ref{rhoI}-\ref{rhoII}).
The last term in Eq.(\ref{rhoI}) was omitted in Eq.(11)  in  Ref.\cite{ZhuPCCP2014}
and the two last terms in Eq.(\ref{rhoII})  were omitted in Eq.(12) in Ref.\cite{ZhuPCCP2014}.
It can be seen that because of the absence of these terms, Eqs. (11-12) of Zhu and Marcus
\cite{ZhuPCCP2014} do not preserve the total probability.

\begin{figure*}[ht]
\includegraphics[width=7 in]{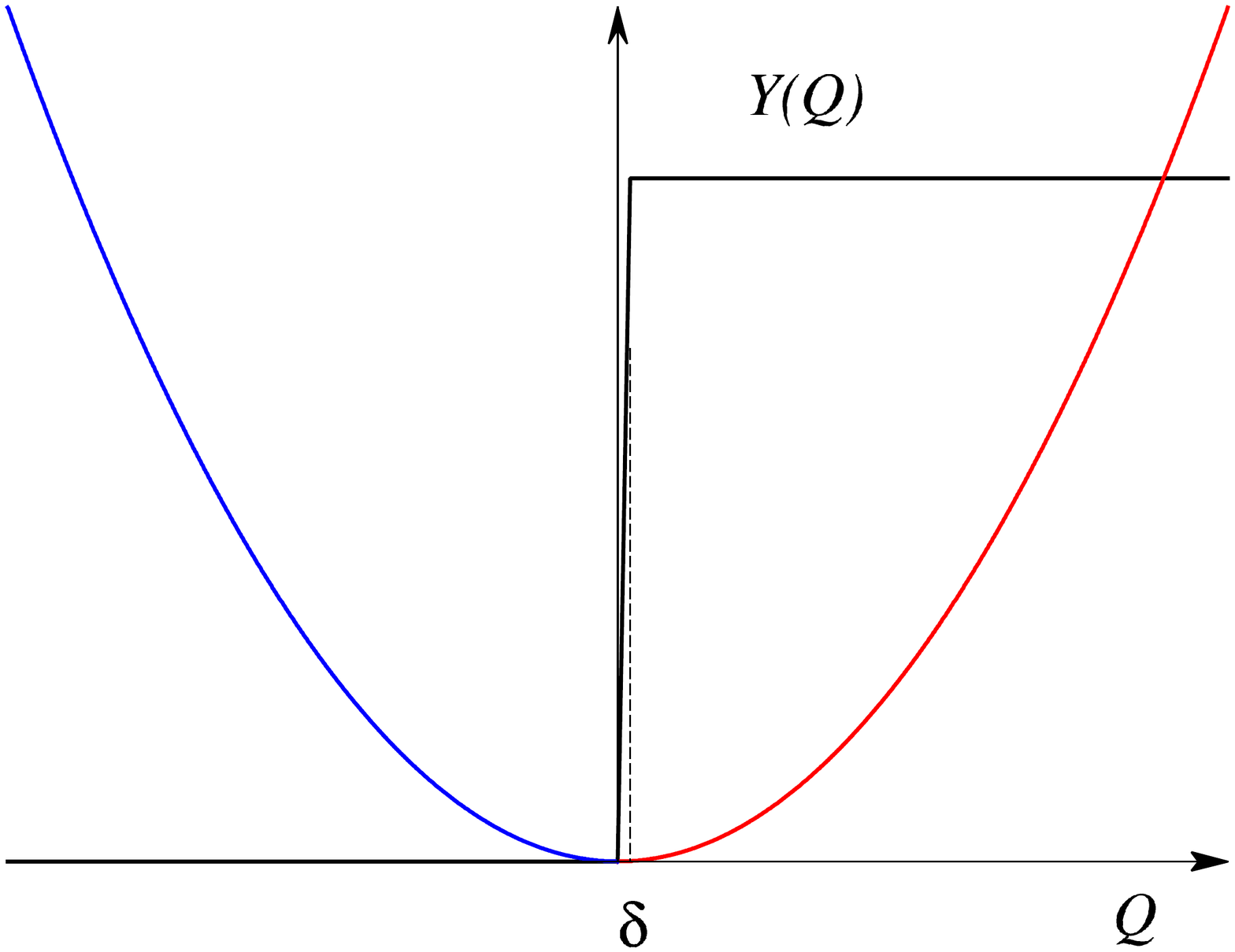}
\caption{The schematic picture of the Frantsuzov and Marcus model.
The potential surface is represented by a line colored red in the bright region ($Q>\delta$) and blue in the dark region ($Q <0$).
The black line represents the PLQY dependence on the coordinate.}
\label{fig:Fran}
\end{figure*}

The ON time and OFF time distribution functions in the Extended  DCET model (\ref{rhoI}-\ref{rhoII}) can be found by solving the following equations:
\be
\frac{\partial}{\partial t}\rho_1(Q,t)=L_1\rho_1(Q,t) -W_1\delta(Q-Q_c)\rho_1(Q,t) -k_{L}\rho_1(Q,t)
\label{rhoI1}
\ee
\be
\frac{\partial}{\partial t}\rho_2(Q,t)=L_2\rho_2(Q,t) -W_2\delta(Q-Q_c)\rho_2(Q,t)
\label{rhoII1}
\ee
Transitions from the dark state to the bright state occur only at the point $Q_c$, thus
the initial distribution for the Eq. (\ref{rhoI1}) is a delta-function:
$$\rho_1(Q,t)=\delta(Q-Q_c)$$
in contrast transitions from a bright state to a dark state can occur not only at the crossing point
and the initial condition for the Eq. (\ref{rhoII1}) has the following form:
$$\rho_2(Q,t)=\int\limits_0^\infty (W_1\delta(Q-Q_c) +k_{L})\rho_1(Q,t)\,dt$$

The Eq.(\ref{rhoI1}) has an additional term  $-k_{L}\rho_1(Q,t)$ in comparison to  Eq.(\ref{EqZi1})
 which leads to an exponential cutoff of the survival probability (\ref{Sur}) time dependence
 $$S_{\mbox {\tiny ON}}(t)=S^0_{\mbox {\tiny ON}}(t)\exp(-k_{L}t)$$
 where $S^0_{\mbox {\tiny ON}}(t)$ is the survival probability obtained from Eq. (\ref{rhoI1}) at $k_{L}=0$.
As a result the ON time distribution function in the Extended DCET model has an exponential cutoff.
$$p_{\mbox {\tiny ON}}(t)\sim \exp(-k_{L}t), \quad   t \gg 1/k_{L} $$
The Eq.(\ref{rhoII1}) is equivalent to  Eq.(\ref{EqZi1}).
The difference in the initial distributions leads to the deviation of the OFF time distribution
in the Extended DCET model in comparison with  the original DCET model at times smaller than $\tau_2$.
The  long time exponential asymptotic behavior, however, is the same
$$p_{\mbox {\tiny OFF}}(t)\sim \exp(-k_{2}t), \quad   t \gg \tau_2 $$
These theoretical predictions are confirmed by numerical simulations (see Fig. \ref{fig:f4}) performed for the case of the symmetric system $Q_c=0$, $W_1=W_2$. The rest of the parameters are    $\tau_1=\tau_2=10^4\,s$, $\Gamma_1=\Gamma_2=10^{-3}\,s^{-1}$, $t_c=t_2=0.1\,s$, $K_{L}=10^{-1}\,s^{-1}$.
It can be concluded that the presence of a second ionization channel resolves the problem with very long ON times, but
not with very long OFF times. As a result, most of the QDs in the Extended DCET model have to become permanently dark
as confirmed by numerical simulations (see Fig. \ref{fig:f5}).
That prediction also significantly differs from the experimentally observed  behavior of single quantum dots.

\section{Frantsuzov and Marcus model}

\begin{figure*}[ht]
\includegraphics[width=7 in]{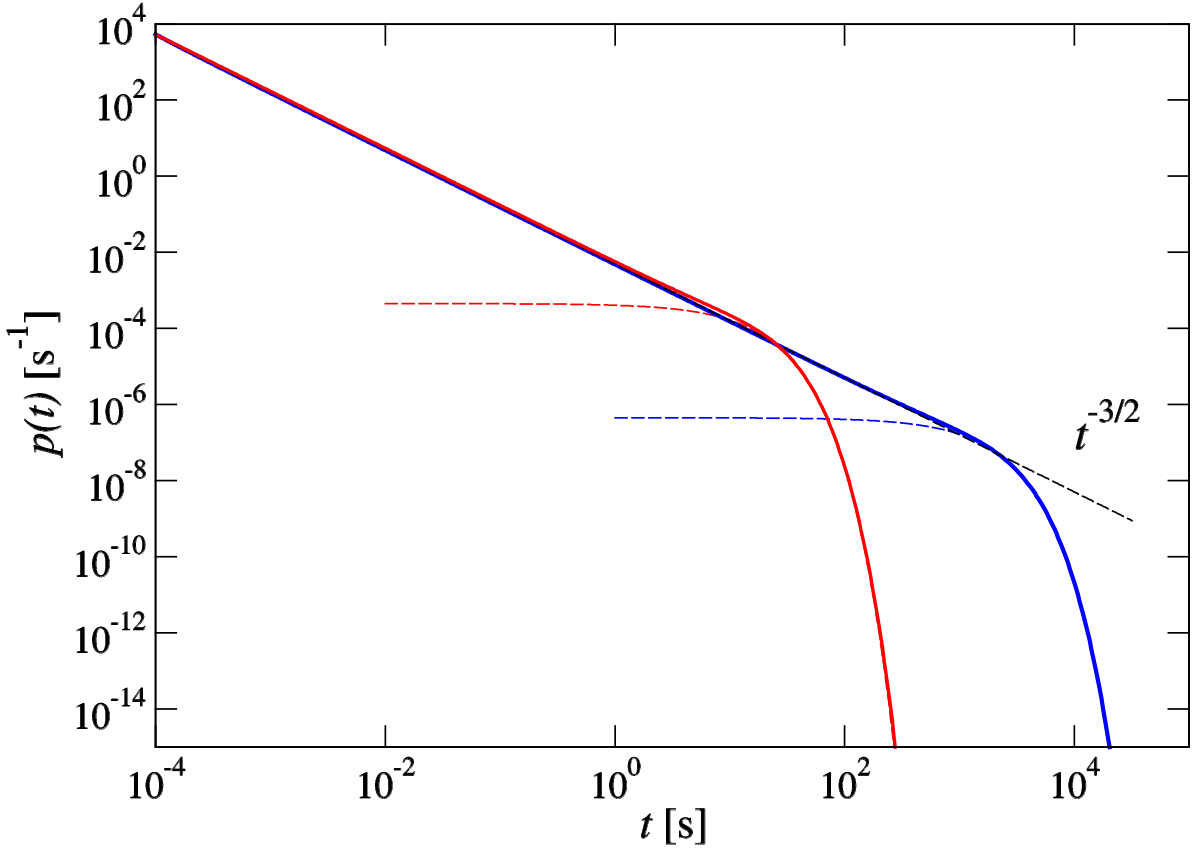}
\caption{The normalized ON time (red thick line)  and OFF time (blue thick line) distributions obtained by numerical simulations in the Frantsuzov and Marcus model,
 the $t^{-3/2}$ dependence  (thin black line), $\exp(-t/T_{\mbox {\tiny ON}} )$ (red dashed line), and $\exp(-t/T_{\mbox {\tiny OFF}} )$ (blue dashed line).
 The parameters of the model are $T_{\mbox {\tiny ON}}=10$ s, $T_{\mbox {\tiny OFF}}=10^3$ s, $\delta=10^{-3}$, $\tau_{m}=10^{-4}$ s}
\label{fig:f6}
\end{figure*}

\begin{figure*}[ht]
\includegraphics[width=7 in]{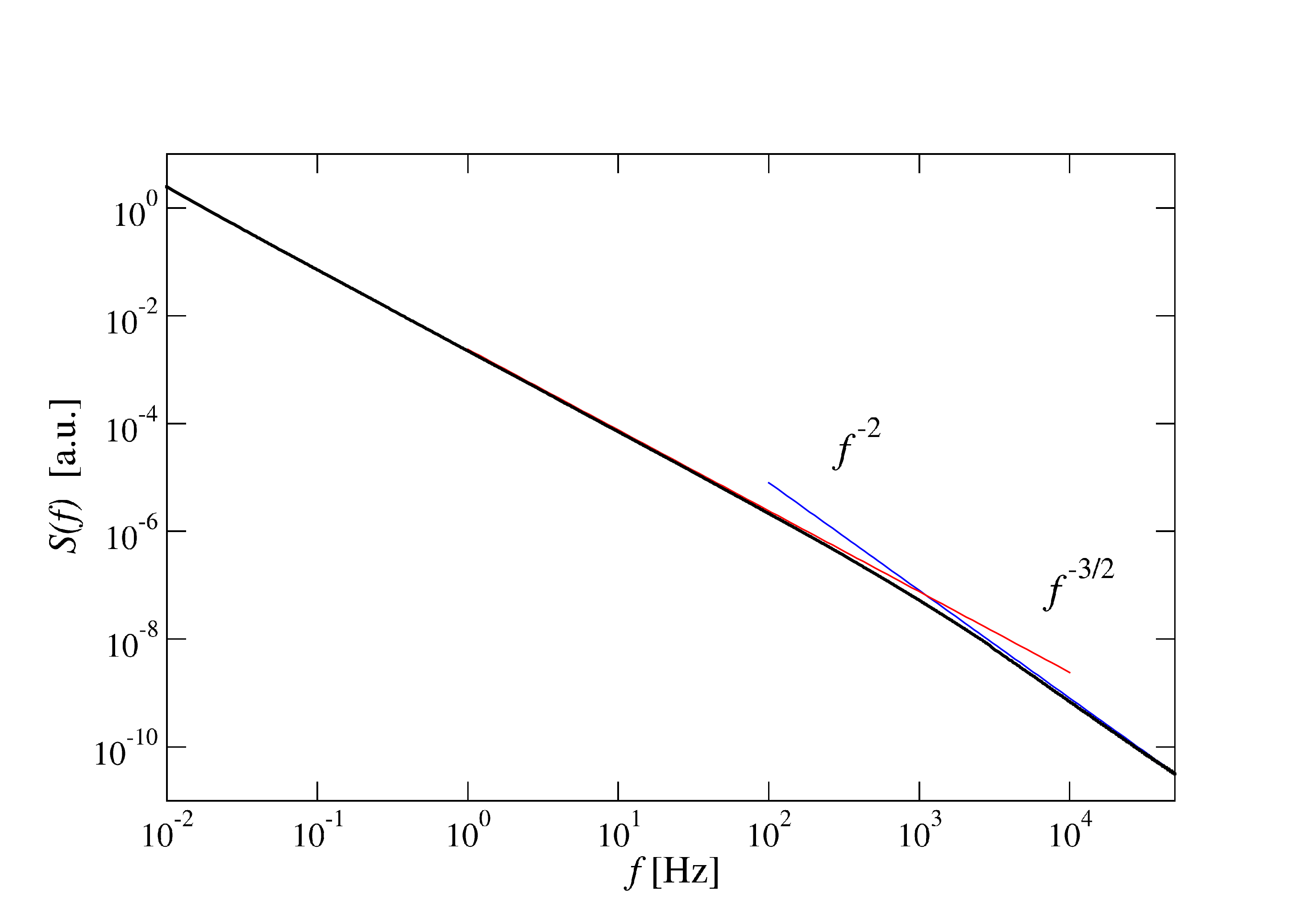}
\caption{Power spectral density of the single QD fluorescence emission quantum yield (thick line) in the  Frantsuzov and Marcus model,
 the $f^{-3/2}$ dependence  (thin red line), and  the $f^{-2}$ dependence  (thin blue line).
Parameters of the model are $T_{\mbox {\tiny ON}}=T_{\mbox {\tiny OFF}}=10^3$ s, $\delta=10^{-3}$.}
\label{fig:f7}
\end{figure*}

The Frantsuzov and Marcus model \cite{FrantsuzovPRB05} is  based on the fluctuating rate mechanism,
thus it does not consider transitions between neutral and charged states.
Fluctuations of the emission intensity in the model are caused by variations
of the PLQY (\ref{Y}).
The nonradiative recombination rate $k_n$ depends on the reaction coordinate $Q$
which is performing diffusive motion.
 Within the generalized formulation of the model
the probability distribution function $\rho(Q,t)$ satisfies the equation
\be
\frac{\partial}{\partial t}\varrho(Q,t)= \frac{\partial}{\partial Q}  D(Q)
\left(\frac{\partial}{\partial Q}+ Q\right) \varrho(Q,t)
\label{rhoQ}
\ee
where $D(Q)$ is the coordinate dependent diffusion coefficient.
To generate fast transitions from high to low emission intensity and back, the function $Y(Q)$ must grow dramatically
from a minimal value to a maximum one on a tiny interval of $\delta$ close to the origin (see Fig. \ref{fig:Fran}).
Thus, the QD is bright when $Q>\delta$, dark when $Q<0$, and has some intermediate florescence intensity within the interval of $\delta\ll 1$.
Taking into account  that a molecular mechanism of the spectral diffusion is light induced  \cite{BawendiJPCB99,MulvaneyPRB10},
 the diffusion coefficient $D(Q)$ has to depend on the excitation intensity.
It also means that the diffusion could be much faster for a bright QD  than for a dark one \cite{FrantsuzovPRB05}.
 As such, we can choose:
 \be
D(Q)= \left\{ \begin{array}{ccl}  1 /{T_{\mbox {\tiny OFF}}}, & & Q<0\\
1/ {T_{\mbox {\tiny ON}}}, &   \delta\le &Q
\end{array} \right.
\label{DQ}
\ee
It was shown by Frantsuzov and Marcus \cite{FrantsuzovPRB05} that the normalized ON time and OFF time distributions obtained by the threshold procedure
have the following dependence (see Appendix E for details):
\be
p(t)=\frac {\sqrt{\tau_{m}}} 2 t^{-3/2},\quad \tau_{m}\le t\ll T_0
\label{pshort}
\ee
\be
p(t)=\sqrt{\frac{2 \tau_{m}}{T_0^3}}  \exp(-t/T_0),\quad  T_0 \ll t
\label{plong}
\ee
where $\tau_m$ is the minimum time interval of observation (bin time) and $T_0$ is equal to $T_{\mbox {\tiny ON}}$ and  $T_{\mbox {\tiny OFF}}$ for the ON time and OFF time distribution, respectively.
That prediction is confirmed by the numerical simulations made using the SSDP program \cite{KrissinelJCC97} (see Fig. \ref{fig:f6}).

The power spectral density  $S(f)$ of the single QD emission
at frequencies $f$  larger than  $1/\tau_m$  could be obtained without binning procedure
by measuring the autocorrelation function  \cite{SionnestAPL04,PeltonPNAS07}.
In order to calculate  $S(f)$ within the model one needs to specify the $Y(Q)$ function
in the intermediate interval. Let's choose the simplest linear dependence:
\be
Y(Q)= \left\{ \begin{array}{ccl} 0, \quad & &Q<0\\
Q/\delta, \quad & 0\le &Q < \delta\\
1,\quad & \delta \le &Q
\end{array} \right.
\label{YQ}
\ee
The results of numerical calculations  of the $S(f)$ in that case are presented in Fig. \ref{fig:f7}
(see Appendix F for the detailed calculation procedure).
The Figure clearly shows the transition from  the $f^{-3/2}$ dependence to  $f^{-2}$ at large frequencies
in accordance with the experiment of Pelton et al. \cite{PeltonPNAS07}.

\section{Discussion}

As a result of the above analytical and numerical studies it was found that two  models of single QD blinking based on spectral diffusion, namely
the DCET model \cite{TangJCP05,TangPRL05} and the Extended DCET model \cite{ZhuPCCP2014} predict that after an initial blinking period, most of the QDs should become permanently bright or  permanently dark. That prediction significantly differs from the behavior of single quantum dots observed in numerous experiments.
Another drawback of these models is the charging mechanism on which they are based.
Despite the fact that most of the theoretical models proposed in the literature are based on that mechanism \cite{KunoJCP01,BawendiPRB01,OrritPRB02,BarkaiJCP04,TangPRL05,OsadkoJPCC13,ZhuPCCP2014},
 there is a number of sufficient experimental evidence indicating that the charging mechanism fails in explaining the QD blinking phenomenon.
In several experiments, the emission intensity of a single QD was observed below the charged state (trion) emission intensity
 \cite{BawendiPRL10,OronPRL10,OronACSNano13,KlimovNL17}.
Another very important set of experiments showed that the existence of the distinct ON and OFF states is an illusion; there is a nearly continuous set of emission intensities \cite{MewsPRL02,BawendiJPCB04,YangNL06,CichosJL11,BorczyskowskiACSNano14}.
Furthermore, it was also shown \cite{FrantsuzovPRL09,PeltonNL10,CichosJCP14} that the parameters $m$ and $T$ of the ON and OFF time distributions strongly depend on the threshold value.

The Frantsuzov and Marcus model \cite{FrantsuzovPRB05}, based on fluctuating rate mechanism, reproduces the key properties of the QD blinking phenomenon.
Nonetheless there are a number of the experimental observations which are not explained by the model:\\
1. The exponent value $m$ of the ON and OFF time distribution functions is reported in the range from 1.2 to 2.0 \cite{FrantsuzovNaturePhys08},
and it strongly depends on the threshold value \cite{FrantsuzovPRL09,PeltonNL10,CichosJCP14}. Meanwhile, in the model, $m$ is always equal to $3/2$ regardless of the threshold.\\
2. The exponent $r$ of the emission power spectral density is found to be in the range from 0.7 to 1.2 \cite{SionnestAPL04,PeltonPNAS07,FrantsuzovNL13}, when
the model predicts the exponent value of 3/2.\\
3. The long-term correlations between subsequent ON and OFF times \cite{StefaniNJP05,VolkanNL10}. There are no such correlations in the model.

A possible reason for this discrepancy is that the description of the spectral diffusion in the model does not fully correspond to its real properties.
It was shown that the squared frequency displacement of the single QD emission has an anomalous (sublinear) time dependence \cite{MulvaneyPRL10}.
Plakhotnik et al. \cite{MulvaneyPRL10} suggested an explanation of  this behavior by introducing a number
of stochastic two-level systems (TLS) having a wide spectrum of flipping rates.
A similar idea was applied by Frantsuzov, Volkan-Kacso and Janko  in the Multiple Recombination Center (MRC) model
of single QD blinking \cite{FrantsuzovPRL09}.
The MRC model, based on the fluctuating rate mechanism, also reproduces  the key properties the single QD blinking.
But in addition it explains the power spectral density dependence close to $1/f$ \cite{FrantsuzovNL13},
the threshold dependence of the $m$ and $T$ values \cite{FrantsuzovPRL09},
 and the long-term correlations between subsequent blinking times \cite{VolkanNL10}.
 This suggests that the spectral diffusion and the fluctuations of the emission intensity of a single QD can be explained by an unified model,
 which could become a generalization of the Frantsuzov and Marcus model.

In conclusion, we analytically and numerically considered three models of the single QD emission fluctuations (blinking) based on spectral diffusion. Only one of them, the Frantsuzov and Marcus model \cite{FrantsuzovPRB05},  reproduces the key properties of the phenomenon.
The DCET model \cite{TangJCP05,TangPRL05} and the Extended DCET model \cite{ZhuPCCP2014} predict that after an initial blinking period, most of the QDs should become permanently bright or  permanently dark which is significantly different from the experimentally observed behavior.

\section*{Acknowledgement}
The authors are very grateful to Professor Rudolph Marcus for fruitful discussions.

The study was supported by the Russian Foundation for Basic Research, project 16-02-00713.

\section*{Appendix A: An analytical solution for the blinking time distribution within the DCET model}

Introducing a dimensionless coordinate $x$
$$x=\frac {Q+E_r} {\sqrt {2E_rkT}}$$
we can rewrite Eq. (\ref{EqZi1}) as
$$\frac{\partial}{\partial t}\rho_1(x,t)=\frac 1 {\tau_1} \frac{\partial}{\partial x}
\left(\frac{\partial}{\partial x}+ x \right) \rho_1(x,t) $$
\begin{equation}
 -W \delta(x-x_c)\rho_{1}(x,t)
\label{Eqx}
\end{equation}
with the initial condition
$$\rho_1(x,0)=\delta(x-x_c)$$
where the relaxation time $\tau_1$ is given by Eq.(\ref{tau1}),
$x_c$ is the dimensionless crossing point coordinate Eq.(\ref{xc})
 and $W$ is given by Eq.(\ref{W})
Applying Eq.(\ref{Eqx}), the ON time distribution function (\ref{pON})  can be expressed as
\be
p_{\mbox {\tiny ON}}(t)=-\frac{d}{dt}\int_{-\infty}^{\infty}  \rho_1(x,t)\,dx=W\rho_1(x_c,t)
\label{pONW}
\ee
The Laplace image of the function $\rho_1(x,t)$
$$\tilde \rho_1(x,s)=\int_0^\infty \rho_1(x,t)e^{-st}\,dt$$
obeys the following equation
$$s\tilde\rho_1(x,s)-\delta(x-x_c)= $$
\begin{equation}
 \frac 1 {\tau_1} \frac{\partial}{\partial x}
\left(\frac{\partial}{\partial x}+ x \right)\tilde \rho_1(x,s)-W \delta(x-x_c)\tilde\rho_{1}(x,s)
\label{EqLap}
\end{equation}
 The Green's function of the differential operator in Eq.(\ref{Eqx})
 satisfies the equation
\begin{equation}
\frac{\partial}{\partial t}G(x,x',t)= \frac 1 {\tau_1} \frac{\partial}{\partial x}
\left(\frac{\partial}{\partial x}+ x \right) G(x,x',t)
\label{GF}
\end{equation}
with the initial condition
$$ G(x,x',0)=\delta(x-x')$$
The Green's function and the Laplace image satisfies the equation
$$s\tilde G(x,x',s)-$$
\begin{equation}
\frac 1 {\tau_1} \frac{\partial}{\partial x}
\left(\frac{\partial}{\partial x}+ x \right) \tilde G(x,x',s)=\delta(x-x')
\label{GFL}
\end{equation}
Using Eq.(\ref{GFL}), Eq.(\ref{EqLap}) can be rewritten as
\be
\tilde\rho_1(x,s)=\tilde G(x,x_c,s)-W\tilde G(x,x_c,s)\tilde\rho_1(x_c,s)
\label{rho1}
\ee
From Eq.(\ref{rho1}) we can find $\tilde\rho_1(x_c,s)$
$$\tilde\rho_1(x_c,s)=\frac {\tilde G(x_c,x_c,s)}{1+W\tilde G(x_c,x_c,s)}$$
The Laplace image of the ON time distribution  Eq.(\ref{pONW}) is given by
$$\tilde p_{\mbox {\tiny ON}}(s)=W\tilde \rho_1(x_c,s)$$
Substituting  Eq.(\ref{EqLap}) we get
\be
\tilde p_{\mbox {\tiny ON}}(s)=\frac {W \tilde G(x_c,x_c,s)}{1+W \tilde G(x_c,x_c,s)}
\label{pONss}
\ee
Green's function (\ref{GF}) is well-known:
\begin{equation}
G(x,x',t)=\frac{1}{\sqrt{2\pi\left(1-e^{-2t/\tau_1}\right)}}
\exp\left[-\frac{\left(x-x' e^{-t/\tau_1}\right)^2}
{2\left(1-e^{-2t/\tau_1}\right)}\right]
\label{Gan}
\end{equation}
Introducing the function $g_1(s)$
\be
g_1(s)=\tilde G(x_c,x_c,s)
\label{g1s}
\ee
we can express Eq.(\ref{pONss})  in the form Eq.(\ref{pONs}).

\section*{Appendix B: The ON time distribution at short times within the DCET model}
At a short time limit $t\ll \tau_1$ one has to find the function $g_1(s)$ at $s \gg 1/\tau_1$.
Expanding the exponent's argument in the Eq. (\ref{gs}) we get
\begin{equation}
g_1(s)=\int_0^\infty \frac{\exp(-st-\Gamma_1 t) }{\sqrt{4\pi t/\tau_1}}\,dt=\frac 1 2 \sqrt{\frac {\tau_1} {s+\Gamma_1}}
\label{gt}
\end{equation}
where $\Gamma_1$ is given by Eq.(\ref{Gamma1})
Substituting  Eq.(\ref{gt}) into Eq.(\ref{pONs}) gives
$$\tilde p_{\mbox {\tiny ON}}(s)= \frac 1 {1+ \sqrt{(s+\Gamma_1)t_c}}$$
and after the inverse Laplace transformation we get Eq. (\ref{pONshort}).

\section*{Appendix C: The ON time distribution at long times within the DCET model}
The approximate formula (\ref{pONshort}) works for short times only.
In order to see the behavior of the function $p_{\mbox {\tiny ON}}(t)$  at a long time limit $t \gg \tau_1$
one has to consider its Laplace image $\tilde p_{\mbox {\tiny ON}}(s)$ (\ref{gs}) at $s\to 0$.
If we expand the function $g_1(s)$ (\ref{g1s}) into a series on $s$:
 \be
g_1(s)\approx \frac 1 s A + B
\label{gexp}
\ee
where
$$A=\lim_{t\to \infty} G(x_c,x_c,t)$$
and
$$B=\int_0^\infty \{G(x_c,x_c,t)-A\} \,dt$$
The Green's function (\ref{Gan}) approaches the stationary distribution at long times
$$\lim_{t\to \infty} G(x,x',t) =\frac{1}{\sqrt{2\pi}} \exp\left(-\frac {x^2}2\right)$$
Thus the constants $A$ and $B$ are
$$A=\frac{1}{\sqrt{2\pi}} \exp\left(-\frac {x_c^2}2\right)$$
 $$B=\int_0^\infty \left[\frac{\exp\left(-\frac 1 2{x_c^2}\tanh\left({\frac{t}{2\tau_1}}\right)\right) }
 {\sqrt{2\pi\left(1-e^{-2t/\tau_1}\right)}}-A\right]\,dt$$

Substituting Eq.(\ref{gexp}) into Eq. (\ref{pONs}) we get the following dependence of $\tilde p_{\mbox {\tiny ON}}(s)$ at small $s$
$$\tilde p_{\mbox {\tiny ON}}(s)\approx \frac {WB}{1+WB}+ \frac  {p_l} {s+k}$$
which corresponds to the exponential behavior Eq. (\ref{Pexp})of the ON time distribution function at long times.

\section*{Appendix D: The derivation of the evolution equations within the Extended DCET model  }
If $k_{eg}$ is much larger than all other rates in Eq. (\ref{Zhu2}) a quasiequlibrium value of exciton population is
established
\be
\varrho_e (Q,t)\approx \frac {I_{ge}} {k_{eg}}  \varrho_d (Q,t)
\label{Zhu9}
\ee
Similarly if $k_{be}$ is much larger than all other rates in Eq.{\ref{Zhu3}:
\be
\varrho_b (Q,t)\approx \frac {I_{eb}} {k_{be}}  \varrho_e (Q,t)
\label{Zhu10}
\ee
If $k_{d'd} \gg k_{bd'}$
\be
\varrho_{d'} (Q,t)\approx\frac {k_{bd'}} {k_{d'd}}  \varrho_b (Q,t)\approx \frac {k_{bd'}} {k_{d'd}} \frac {I_{eb}} {k_{be}}  \varrho_e (Q,t)
\label{Zhu11}
\ee
Substituting Eqs. (\ref{Zhu9}-\ref{Zhu11}) with the definition Eq. (\ref{Zhu7}) into Eqs. (\ref{Zhu1}-\ref{Zhu3}) we get
Eq.(\ref{rhoI}).

If $k_{d^\ast d}$ is much larger than all other rates in Eq. (\ref{Zhu5}) a quasiequlibrium value of the dark exciton population is
established
\be
\varrho_{d^\ast} (Q,t)=\frac {I_{ge}} {k_{d^\ast d}}   \varrho_d (Q,t)
\label{Zhu12}
\ee
Substituting Eq.(\ref{Zhu12}) with the definition Eq.(\ref{Zhu8})  into Eqs.(\ref{Zhu4}-\ref{Zhu6}) we
obtain Eqs.(\ref{rhoII}).

\section*{Appendix E: The ON time  and OFF time distributions within the Frantsuzov and Marcus model}
 The survival probability of the ON time within the Frantsuzov and Marcus model can be found as an integral
 \be
S_{\mbox {\tiny ON}}(t)=\int\limits_0^\infty \rho(Q,t)\,dQ
\label{SurFr}
\ee
 where   $\rho(Q,t)$ is a solution of the following equation
 \be
\frac{\partial}{\partial t}\rho(Q,t)= \frac 1 {T_{\mbox {\tiny ON}}} \frac{\partial}{\partial Q}
\left(\frac{\partial}{\partial Q}+ Q\right) \rho(Q,t)
\label{rhoon}
\ee
 with an absorbing boundary condition at the border (the first passage time problem)
 \be
\left.\rho(Q,t)\right|_{Q=0} = 0
\label{bound}
\ee
The question of what to take as the initial distribution for the equation is not easily answered.
There is the minimal time $\tau_m$ (bin time) of the ON time  period which can be observed.
In accordance with Eq.(\ref{rhoon}),  if the ON time period is longer than $\tau_m$  then the coordinate $Q$ has reached values larger than
  $\sqrt{\tau_m/T_{\mbox {\tiny ON}}}$.
We can take any distribution located at a distance less than  $\sqrt{\tau_m/T_{\mbox {\tiny ON}}}$ from the origin as an initial one.
For the sake of simplicity, we can take the initial distribution in the form of a delta function
\be
\rho(Q,0)=\delta(Q-\Delta)
\label{init}
\ee
where
$$\delta \ll \Delta \ll \sqrt{\tau_m/T_{\mbox {\tiny ON}}}$$

The solution of Eqs.(\ref{rhoon}-\ref{init}) is well known
\be
\rho(Q,t)=G(Q,\Delta,t)-G(-Q,\Delta,t)
\label{rhoF}
\ee
where $G(x,x',t)$ is the Green's function of the Eq.(\ref{rhoon})
\begin{equation}
G(Q,Q',t)=\frac{
\exp\left\{-\frac{\left[Q-Q' \exp(-t/T_{\mbox {\tiny ON}})\right]^2}
{2\left(1-\exp(-2t/T_{\mbox {\tiny ON}})\right)}\right\}}
{\sqrt{2\pi\left(1-\exp(-2t/{T_{\mbox {\tiny ON}}})\right)}}
\label{GFR}
\end{equation}
Using Eq.(\ref{rhoF}) the survival probability (\ref{SurFr}) can be expressed as
$$ S_{\mbox {\tiny ON}}(t)=
\frac { \int\limits_{-b}^b
 \exp\left[-\frac{Q^2}
{2\left(1-\exp(-2t/T_{\mbox {\tiny ON}})\right)}\right]\,dQ}
{\sqrt{2\pi\left(1-\exp(-2t/T_{\mbox {\tiny ON}})\right)}}$$
where $b=\Delta \exp(-t/T_{\mbox {\tiny ON}})$.  At times $t>\tau_m$  the expression
can be rewritten as
 $$S_{\mbox {\tiny ON}}(t)=\frac{2\Delta \exp(-t/T_{\mbox {\tiny ON}})}{\sqrt{2\pi\left(1-\exp(-2t/T_{\mbox {\tiny ON}})\right)}}$$
This expression has the following behavior in the limiting cases
\be
S_{\mbox {\tiny ON}}(t)=  \Delta \sqrt{\frac {T_{\mbox {\tiny ON}}}{\pi t}}, \quad  \Delta^2 T_{\mbox {\tiny ON}} \ll t \ll T_{\mbox {\tiny ON}}
\label{Sshort}
\ee
\be
S_{\mbox {\tiny ON}}(t)=  \Delta \sqrt{\frac 2 \pi}\exp(-t/T_{\mbox {\tiny ON}}), \quad    T_{\mbox {\tiny ON}} \ll t
\label{Slong}
\ee

In the experiment one can see that only the ON times are longer than $\tau_m$, which means that the ON time distribution should be normalized as follows
$$\int\limits_{\tau_m}^\infty p_{\mbox {\tiny ON}}(t)\,dt=1$$
The normalization procedure is equivalent to scaling of the function $S_{\mbox {\tiny ON}}$ so that the following equality for the normalized
survival probability is satisfied
\be
\bar S_{\mbox {\tiny ON}}(\tau_m)=1
\label{Stau}
\ee
Applying this normalization to Eqs. (\ref{Sshort}-\ref{Slong}) we get
$$
\bar S_{\mbox {\tiny ON}}(t)=  \sqrt{\frac {\tau_m}{t}}, \quad  \tau_m \le t \ll T_{\mbox {\tiny ON}}$$
$$
\bar S_{\mbox {\tiny ON}}(t)=  \sqrt{\frac 2 {T_{\mbox {\tiny ON}} }}\exp(-t/T_{\mbox {\tiny ON}}), \quad    T_{\mbox {\tiny ON}} \ll t $$

From Eq.(\ref{pON}) we obtain the ON time distribution function
$$
p_{\mbox {\tiny ON}}(t)=  \frac 1 2  \sqrt{\tau_m}t^{-3/2}, \quad  \tau_m \le t \ll T_{\mbox {\tiny ON}}$$
$$
p_{\mbox {\tiny ON}}(t)=   \sqrt{\frac {2 \tau_m} {T_{\mbox {\tiny ON}}^3}} \exp(-t/T_{\mbox {\tiny ON}}), \quad    T_{\mbox {\tiny ON}} \ll t$$
Similarly the expression for the OFF time distribution function can be obtained
$$
p_{\mbox {\tiny OFF}}(t)=  \frac 1 2  \sqrt{\tau_m}t^{-3/2}, \quad  \tau_m \le t \ll T_{\mbox {\tiny OFF}}$$
$$
p_{\mbox {\tiny ON}}(t)=   \sqrt{\frac {2 \tau_m} {T_{\mbox {\tiny OFF}}^3}} \exp(-t/T_{\mbox {\tiny OFF}}), \quad    T_{\mbox {\tiny OFF}} \ll t$$
These expression are equivalent to the Eqs. (\ref{pshort}-\ref{plong}).

\section*{Appendix F: The emission intensity autocorrelation function within the Frantsuzov and Marcus model}
The autocorrelation function of the emission intensity within the FRM is
$$C(t)=\left\langle Y\left(Q(t)\right)Y\left(Q(0)\right) \right\rangle$$
where averaging is performed  over the ensemble of realizations of the random process $Q(t)$.
For the Frantsuzov and Marcus model the function $C(t)$  can be written as
\be
C(t)=\int\limits_{-\infty}^\infty \int\limits_{-\infty}^\infty Y(Q) G(Q,Q',t) Y(Q')\varrho_0(Q')\,dQdQ'
\label{Ct}
\ee
where $G(Q,Q',t)$ is the Green's function  of Eq. (\ref{rhoQ})  and the stationary distribution  $\varrho_0$ is
$$\varrho_0(Q)=\frac 1 {\sqrt{2\pi}} \exp(-\frac 1 2 Q^2)$$
Eq. (\ref{Ct}) can be rewritten as
$$C(t)=\int\limits_{-\infty}^\infty Y(Q) \varrho(Q,t)\,dQ$$
where $\varrho(Q,t)$ is the solution of Eq. (\ref{rhoQ})
with the initial condition
$$\varrho(Q,0)=Y(Q)\varrho_0(Q)$$
A numerical solution was obtained using the SSDP program \cite{KrissinelJCC97}.
The power spectral density was calculated using a cosine transform
$$S(f)=4\int\limits_0^\infty C(t)\cos(2 \pi f t)\,dt $$

\bibliography{FrantsuzovOS}

\end{document}